\documentclass[twocolumn]{aastex63}
\usepackage{natbib}
\newcommand{\squiggle}{SQuIGG$\vec{L}$E \,}

\newcommand{\medmassdiffq}{.0010 dex\,}
\newcommand{\medmassdiffsf}{.0010 dex\,}
\newcommand{\medzdiffq}{$0.006$\,}
\newcommand{\medzdiffsf}{$0.007$\,}

\newcommand{\tqmedian}{$154$ Myr\,}
\newcommand{\youngdistfraction}{$\sim70\%$} %{$70.9\%$}% \pm13.5\%$\,}
\newcommand{\olddistfraction}{$\sim40\%$}%{$39.3\%$}% \pm13.4\%$\,}

\usepackage{amsmath} % or simply amstext

\usepackage{totcount}
\newtotcounter{citnum} %From the package documentation
\def\oldbibitem{} \let\oldbibitem=\bibitem
\def\bibitem{\stepcounter{citnum}\oldbibitem}

\usepackage{float}
\usepackage{textcomp}
\usepackage{makecell}
\usepackage{hyperref}

\date{\today}

\submitjournal{ApJ}

\shorttitle{Mergers \squiggle}

\begin{document}

\title{Merger Signatures are Common, but not Universal, in Massive, Recently-Quenched Galaxies at $z\sim0.7$}

\author[0000-0003-1535-4277]{Margaret Verrico}
\affiliation{Department of Physics and Astronomy and PITT PACC, University of Pittsburgh, Pittsburgh, PA 15260, USA}
\affiliation{Department of Astronomy, University of Illinois, 1002 W. Green St., Urbana, IL 61801, USA}

\author[0000-0003-4075-7393]{David J. Setton}
\affiliation{Department of Physics and Astronomy and PITT PACC, University of Pittsburgh, Pittsburgh, PA 15260, USA}

\author[0000-0001-5063-8254]{Rachel Bezanson}
\affiliation{Department of Physics and Astronomy and PITT PACC, University of Pittsburgh, Pittsburgh, PA 15260, USA}

\author[0000-0002-5612-3427]{Jenny E. Greene}
\affiliation{Department of Astrophysical Sciences, Princeton University, Princeton, NJ 08544, USA}

\author[0000-0002-1714-1905]{Katherine A. Suess}
\affiliation{Astronomy Department, University of California, Berkeley, CA 94720, USA}
\affiliation{Department of Astronomy and Astrophysics, University of California, Santa Cruz, 1156 High Street, Santa Cruz, CA 95064 USA}
\affiliation{Kavli Institute for Particle Astrophysics and Cosmology and Department of Physics, Stanford University, Stanford, CA 94305, USA}

\author[0000-0003-4700-663X]{Andy D. Goulding}
\affiliation{Department of Astrophysical Sciences, Princeton University, Princeton, NJ 08544, USA}

\author[0000-0003-3256-5615]{Justin S. Spilker}
\affiliation{Department of Physics and Astronomy and George P. and Cynthia Woods Mitchell Institute for Fundamental Physics and Astronomy, Texas A\&M University, 4242 TAMU, College Station, TX 77843-4242}

\author[0000-0002-7613-9872]{Mariska Kriek} 
\affiliation{Leiden Observatory, Leiden University, P.O.Box 9513, NL-2300 AA Leiden, The Netherlands}

\author[0000-0002-1109-1919]{Robert Feldmann}
\affiliation{Institute for Computational Science, University of Zurich, CH-8057 Zurich, Switzerland}

\author[0000-0002-7064-4309]{Desika Narayanan}
\affiliation{Department of Astronomy, University of Florida, 211 Bryant Space Science Center, Gainesville, FL, 32611, USA}
\affiliation{University of Florida Informatics Institute, 432 Newell Drive, CISE Bldg E251 Gainesville, FL, 32611, US}
\affiliation{Cosmic Dawn Centre at the Niels Bohr Institute, University of Copenhagen and DTU-Space, Technical University of Denmark}

\author[0000-0002-1759-6205]{Vincenzo Donofrio}
\affiliation{Department of Physics and Astronomy and George P. and Cynthia Woods Mitchell Institute for Fundamental Physics and Astronomy, Texas A\&M University, 4242 TAMU, College Station, TX 77843-4242}

\author[0000-0002-3475-7648]{Gourav Khullar}
\affiliation{Department of Physics and Astronomy and PITT PACC, University of Pittsburgh, Pittsburgh, PA 15260, USA}

\correspondingauthor{Margaret Verrico}
\email{MEV41@pitt.edu}

\begin{abstract}

We present visual classifications of merger-induced tidal disturbances in 143 $\rm{M}_* \sim 10^{11}\rm{M}_\odot$ post-starburst galaxies at z$\sim$0.7 identified in the \squiggle Sample. This sample spectroscopically selects galaxies from the Sloan Digital Sky Survey that have stopped their primary epoch of star formation within the past $\sim$500 Myrs. Visual classifications are performed on Hyper Suprime Cam (HSC) imaging. We compare to a control sample of mass- and redshift-matched star-forming and quiescent galaxies from the Large Early Galaxy Census and find that post-starburst galaxies are more likely to be classified as disturbed than either category.  This corresponds to a factor of $3.6^{+2.9}_{-1.3}$ times the disturbance rate of older quiescent galaxies and $2.1^{+1.9}_{-.73}$ times the disturbance rate of star-forming galaxies. Assuming tidal features persist for $\lesssim500$ Myr, this suggests merging is coincident with quenching in a significant fraction of these post-starbursts. Galaxies with tidal disturbances are younger on average than undisturbed post-starburst galaxies in our sample, suggesting tidal features from a major merger may have faded over time.  This may be exacerbated by the fact that, on average, the undisturbed subset is fainter, rendering low surface brightness tidal features harder to identify. However, the presence of ten young ($\lesssim150$ Myr since quenching) undisturbed galaxies suggests that major mergers are not the only fast physical mechanism that shut down the primary epoch of star formation in massive galaxies at intermediate redshift.

\end{abstract}

\keywords{Post-starburst galaxies (2176), Galaxy quenching (2040), Galaxy evolution (594), Quenched galaxies (2016), Galaxies (573), Galaxy mergers(608)}

%%%%%%%%%%%%%%%%%%%%%%%%%%%%%%%%%%%%%%%%%%%%%%%%%%%%%%%%%%%%%%%%%%%%%%%%%%%%%%%%%%%%%%%%%%%%%%%%%%%%%%%%%%%%%%%%%%%%%%%%%%%%%%%%%%%%%%%%%%%%%%%%%%%%%%%%%%%%%%%%%%%%%%%%%%%%%%%%%%%%%%%%%%%%%%%%%%%%%%%%%%%%%%%%%%%%%%%%%%%%%%%%%%%%%%%%%%%%%%%%%%%%%%%%%%

\section{Introduction} \label{sec:intro}

\begin{figure*}[t!]
\includegraphics[width=0.48\textwidth]{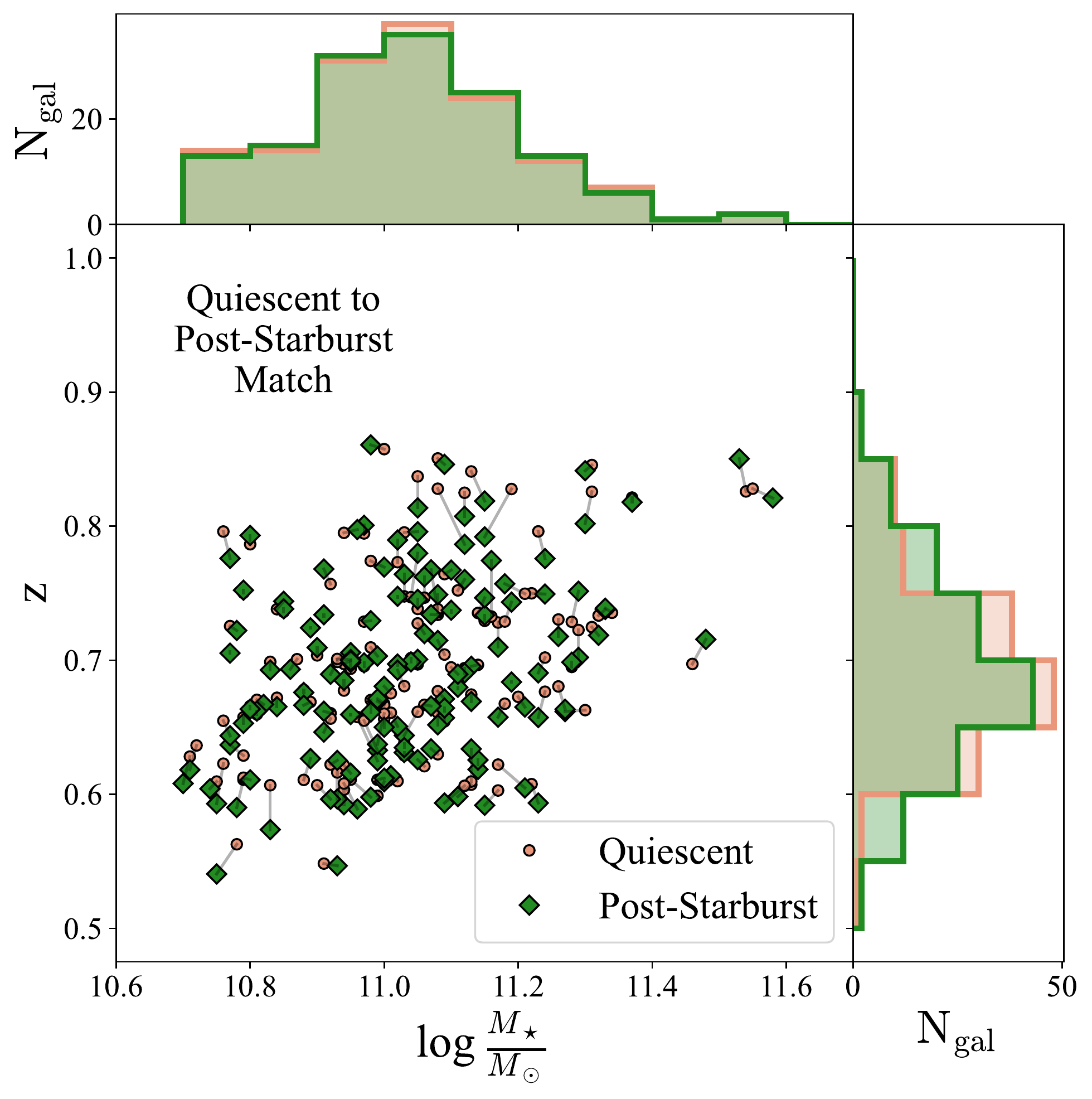}
\includegraphics[width=0.48\textwidth]{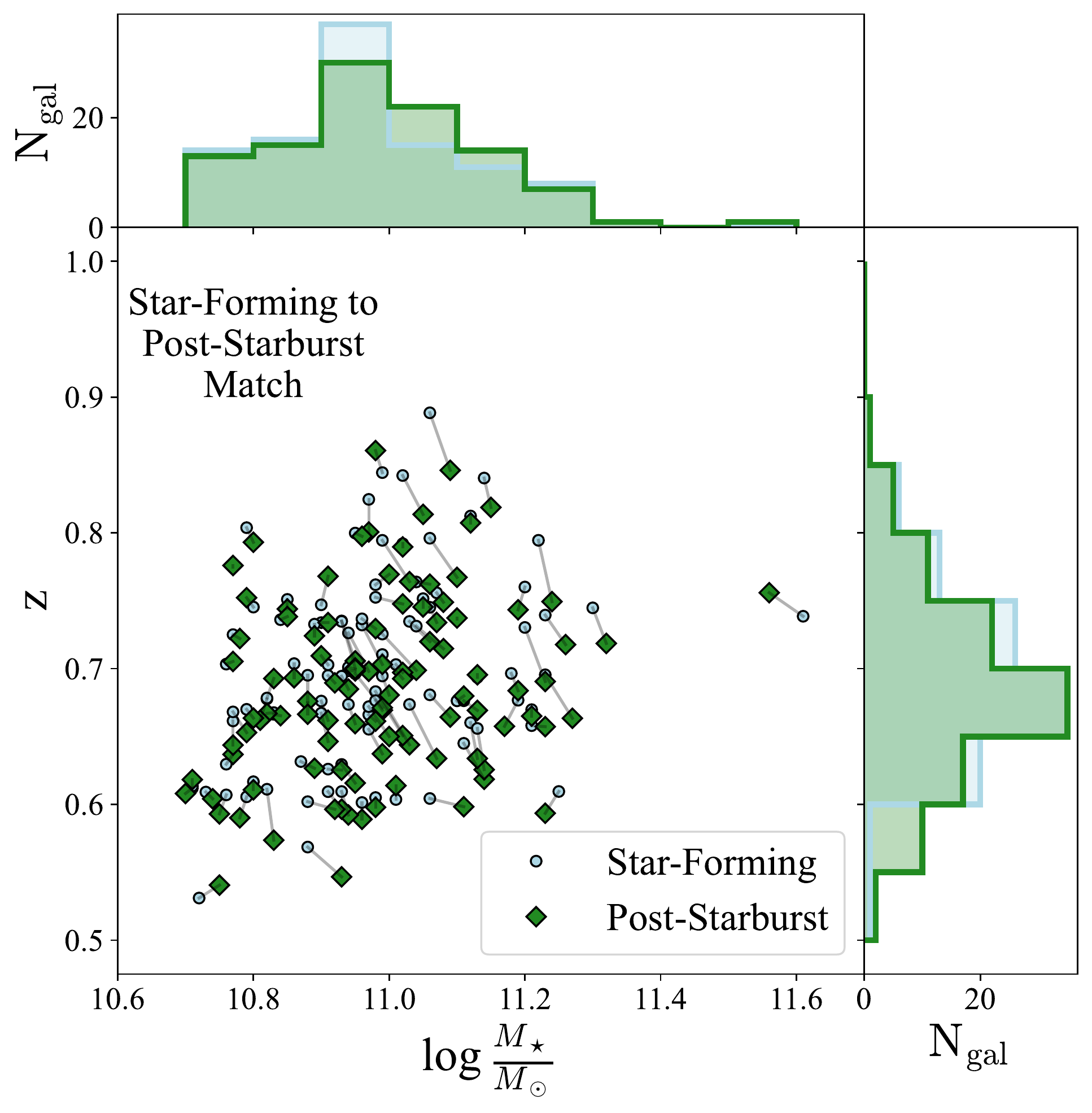}
\caption{The stellar mass versus the redshift of \squiggle post-starburst and quiescent (left) and star-forming (right) LEGA-C control samples. To construct our comparison samples, we first define a Euclidean distance between galaxies in log mass-redshift space.  We then minimize the cumulative distance in mass-redshift space between the post-starburst sample and each of the two control samples.  After removing matches which exceed a difference of 0.05 dex in mass or 0.05 in redshift, we are able to compare 143 post-starburst galaxies to 143 quiescent galaxies and 103 post-starburst galaxies to 103 star-forming galaxies. \label{fig:mass_matching}}
\end{figure*}

Galaxies in the local universe can be broadly divided into two categories: disky galaxies with high star formation rates and spiral arms, and elliptical galaxies which have quenched their star formation and are now ``red and dead." The processes by which galaxies shut off star formation and structurally transition are an area of active study. Recent evidence suggests that the shutdown of star formation can be rapid or drawn out in time, sometimes referred to as a fast and slow mode \citep[e.g.,][]{Barro2013, Schawinski2014, Wild2016, Wu2018, Belli2019, Suess2021}. At low redshift, the slow process dominates, and galaxies quench by exhausting the gas that they use to make stars, while observations of the spectra of quenched galaxies at high redshift suggests that the fast mode dominated at earlier times \citep[e.g.,][]{Wu2018, Belli2019, Suess2021}. The exact physical drivers of this fast mode are still unclear; however, it appears that both the star formation rate and the morphology of these galaxies must ultimately change. There exist several proposed pathways to rapidly quench star formation in galaxies, including compaction due to violent disk instabilities \citep[e.g.,][]{Franx2008, Dekel2013, Ceverino2015} and major galaxy mergers \citep[e.g.,][]{Bekki2005, Hopkins2008, Wellons2015}.

The merging of two or more galaxies can explain a change in both star formation rate and morphology. Gas-rich mergers can distort the ordered disks of star-forming galaxies and pull them apart into irregularly-shaped galaxies with bridges and tidal tails \citep{Toomre1972}. Mergers can funnel gas to the centers of galaxies, making them more compact and suppressing future star formation \citep{Wellons2015,Zheng2020,Pathak2021}.  Mergers play a crucial role in galaxy evolution in the early universe, with the average massive galaxy experiencing more than four major mergers at $z>1$ \citep{Conselice2006}. Observationally, the disrupted morphologies induced by mergers can be identified through visual classification, \citep[e.g.,][]{Lintott2008,Lintott2011,Kartaltepe2015}, non-parametric structural measures \citep[e.g.,][]{Lotz2008, Pawlik2016, Kado-Fong2018, Kim2021}, or even neural networks \citep[e.g.,][]{Bickley2021}. Clear merger signatures are particularly common in ultra-luminous infrared galaxies (ULIRGs) \citep{Goto2005,Kartaltepe2012}, luminous active galactic nuclei (AGN) \citep{Goulding2018}, and recently-quenched galaxies \citep{Lotz2008,Brown2009,Pawlik2016,Yang2008,Pracy2009}, indicating that mergers may play a role in the enhancement of star-formation and the subsequent feedback that could quench star formation in massive galaxies.  

The simultaneity of major mergers and rapid quenching can be tested by looking for merger features in galaxies immediately after they experience a rapid shutdown in star formation. For this reason, we study post-starburst galaxies, a type of galaxy which has recently undergone a period of significant star formation that has rapidly ceased \citep[for a detailed review of post-starburst galaxies, see][]{French2021}. As transitional galaxies, post-starburst galaxies are an excellent test-bed for studying the physical drivers of rapid quenching. A number of studies have used structures of post-starburst galaxies to test this simultaneity. Local studies have found that post-starburst galaxies are often irregular, with 23 to 88\% showing evidence of tidal disturbance that may indicate a recent major merger \citep{Zabludoff1996,Yang2008,Brown2009,Pracy2009,Pawlik2016,Sazonova2021,Wilkinson2022}. Massive post-starburst galaxies are rare but increase in number density with redshift \citep{Whitaker2012a, Wild2016}. In this work, we push morphological study of this critical population to earlier times, inching closer to the era when this phase represented a more dominant occurrence for the population of massive galaxies. 

Throughout this paper we quote AB magnitudes and adopt a concordance $\Lambda$CDM cosmology with $\Omega_{\Lambda}=0.7$, $\Omega_m=0.3$ and $H_0=70$ $\mathrm{km\,s^{-1}\,Mpc^{-1}}$. We utilize deep images of post-starburst galaxies at $z\sim0.7$ to assess the role of mergers in quenching post-starburst galaxies and to supplement our structural analysis of these galaxies in \citealt{Setton2022}.

\section{Data} \label{sec:data}

The \squiggle Sample of post-starburst galaxies was selected from the Sloan Digital Sky Survey (SDSS) DR14 spectroscopic sample \citep{Abolfathi2018} to identify bright galaxies at intermediate-redshift ($z>0.5$) which have recently quenched their primary epoch of star formation using the selection techniques described in \citealt{Suess2022a}. In short, the galaxies were selected with two rest-frame color cuts to isolate the Balmer break, identifying galaxies with strong breaks and blue slopes redward of the break, indicating a dominant ``A" type star population that formed in the past $\sim500$ Myr. 

The sample consists of 1318 unique galaxies, each with SDSS spectroscopy and photometry. However, the faint and compact galaxies are largely unresolved in SDSS imaging, making classification of structures and tidal features impossible. For this work, we utilize the subsample of \squiggle galaxies which overlap with the Hyper-Suprime Cam Survey PDR 3 \citep{Aihara2018, Aihara2021}. For the purposes of this study, we restrict our analysis to the $i$-band imaging, which is observed under the best seeing conditions (PSF FWHM $\sim0.6$") to ensure the highest possible resolution imaging for identification of tidal features (limiting surface brightness for identifying tidal features $\sim26.4$ mag/arcsecond$^2$, see \citealt{Kado-Fong2018}). After visually inspecting and removing galaxies with imaging artifacts, 145 post-starburst galaxies remain with HSC Wide imaging. 

\begin{figure*}[t!]
\includegraphics[width=\textwidth]{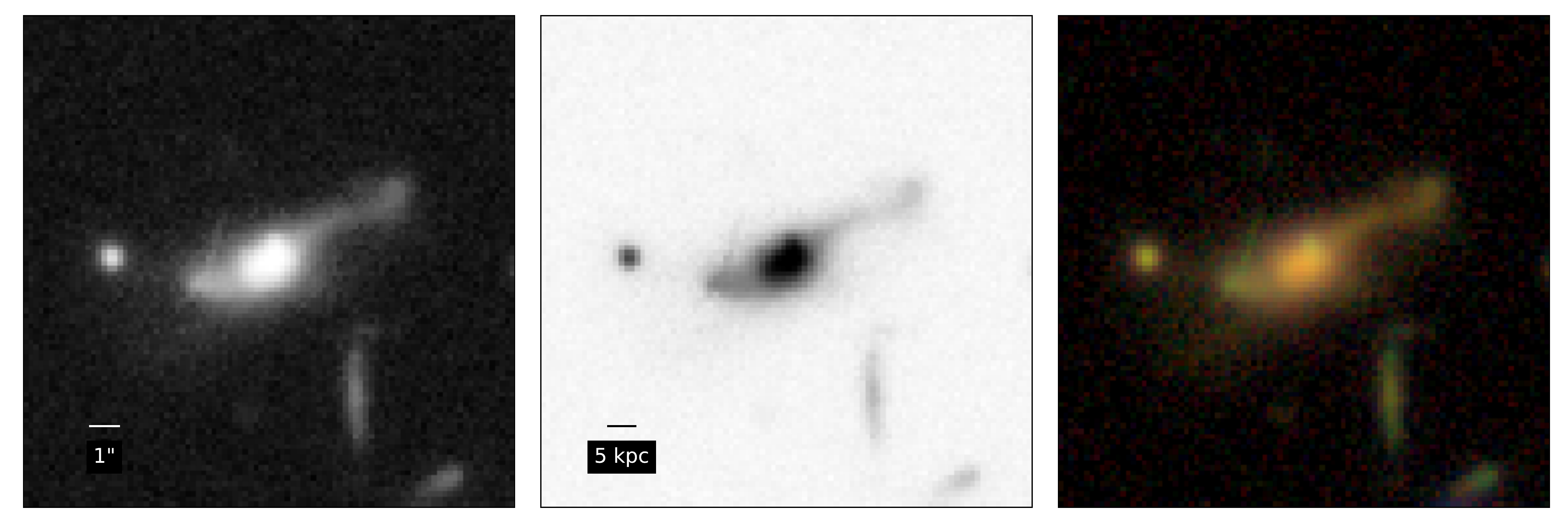}
\caption{An example of the images given to classifiers to perform a visual classification.  The leftmost galaxy is $i$-band imaging plotted white-on-black, the center galaxy is $i$-band imaging plotted black-on-white, and the rightmost galaxy is a 3-color $riz$ image.  For galaxies lacking $r$ or $z$ band imaging, we did not provide a color image.
\label{fig:example}}
\end{figure*}

To compare to co-eval quiescent and star-forming galaxies, we turn to the Large Early Galaxy Census (LEGA-C) Survey DR3 \citep{VanDerWel2021}. The LEGA-C Survey deep (${\sim}20$ hours/galaxy) spectroscopic survey of ${\sim}4000$ galaxies at $0.6<z<1.0$ in the COSMOS field is K-band selected to be mass-representative above ${\sim}10^{10} \ M_\odot$. We use U-V and V-J rest-frame colors to robustly classify these galaxies into star-forming and quiescent populations following the technique described in \citealt{Muzzin2013}. Specifically, galaxies are classified as quiescent if (1) $U-V > 1.3, V-J < 1.5$ and (2) $U-V > (V-J) \times 0.88 + 0.69$. The majority of the LEGA-C star-forming galaxies fall on the star-forming main sequence given in \cite{Whitaker2012b}.  Considering this and that our control sample is at an intermediate redshift, it is unlikely that the star-forming control sample contains starburst galaxies which may be more likely to be merger remnants. However, it is possible that our star-forming control sample is contaminated with some starburst galaxies, and as such, it is possible that fewer true star-forming galaxies are disturbed than we classify in this work.  Crucially, the galaxies in LEGA-C all overlap with the HSC footprint in the COSMOS ultra-deep area, and we utilize images at the HSC-Wide depth for consistency with this analysis \citep[see][]{Aihara2018}.  There is no overlap between SQuIGGLE and LEGA-C samples, as no \squiggle galaxies fall in the COSMOS field.

The LEGA-C sample was selected using an evolving K-band magnitude limit, which was designed to select an approximately stellar mass-representative ($\log M_*/M_{\odot} \gtrsim 10.3$) sample of galaxies \citep[for more detail, see][]{VanDerWel2016}. The \squiggle selection is more complicated due to its stringent spectroscopic signal-to-noise cut on top of the SDSS target selections (most \squiggle galaxies come from bright, red sequence-selected BOSS). In addition, the LEGA-C survey has a sky area of approximately 1.6 square degrees \citep{VanDerWel2016}, in contrast with the HSC Wide survey that covers about 1400 square degrees \citep{Aihara2018}. Thus \squiggle is able to identify intrinsically more rare galaxies, which is useful in observing the brief post-starburst phase but skews masses to the very high end of the mass function. As a result, the mass and redshift distributions of LEGA-C and \squiggle are quite different. To mitigate this effect, we choose to define a Euclidean distance in mass-redshift space and match each \squiggle galaxy to a LEGA-C star-forming and quiescent galaxy such that the cumulative Euclidean distance between the samples is minimized. We then impose a limit in redshift distance of 0.05 and a limit in stellar mass difference of 0.05 dex and remove any matches outside this limit from the sample. We also remove any control images that contain a foreground object and would therefore make visual identification of the central object impossible. Note that this differs from the method used in the companion paper \cite{Setton2022}, in which galaxies were first matched in mass and later in redshift, resulting in a larger number of comparison galaxies. After mass- and redshift-matching, we are able to compare 143 post-starburst and quiescent galaxy pairs and 103 post-starburst and star-forming galaxy pairs.

The results of this matching are shown in Figure \ref{fig:mass_matching}. We are able to match all but two of the \squiggle galaxies to a quiescent counterpart in LEGA-C, but due to the lower masses of star-forming galaxies at this redshift, we are left with only 103 matches to star-forming galaxies. Both samples agree well, with a median mass difference of \medmassdiffq between post-starburst and quiescent galaxies and \medmassdiffsf between post-starburst and star-forming galaxies, and a median redshift difference of \medzdiffq between post-starburst and quiescent galaxies and \medzdiffsf between post-starburst and star-forming galaxies.

\section{Merger Incidence in Post-Starburst Galaxies}

Our goal is to identify significant tidal features that indicate the presence of a major galaxy merger immediately preceding or coincident with quenching. In order to quantify the merger fraction in each sample, we must cleanly distinguish between tidally disturbed and undisturbed galaxies. Quantitative measurements of disturbance, such as asymmetry, shape asymmetry, and Gini-M20, are generally designed to replicate the results of visual classification for large galaxy samples and must be calibrated using simulations and specific imaging characteristics \citep[e.g.][]{Lotz2008,Pawlik2016}. Our sample is small enough that we elect to classify our galaxies visually. 

\begin{figure*}[t!]
\includegraphics[width=0.5\textwidth]{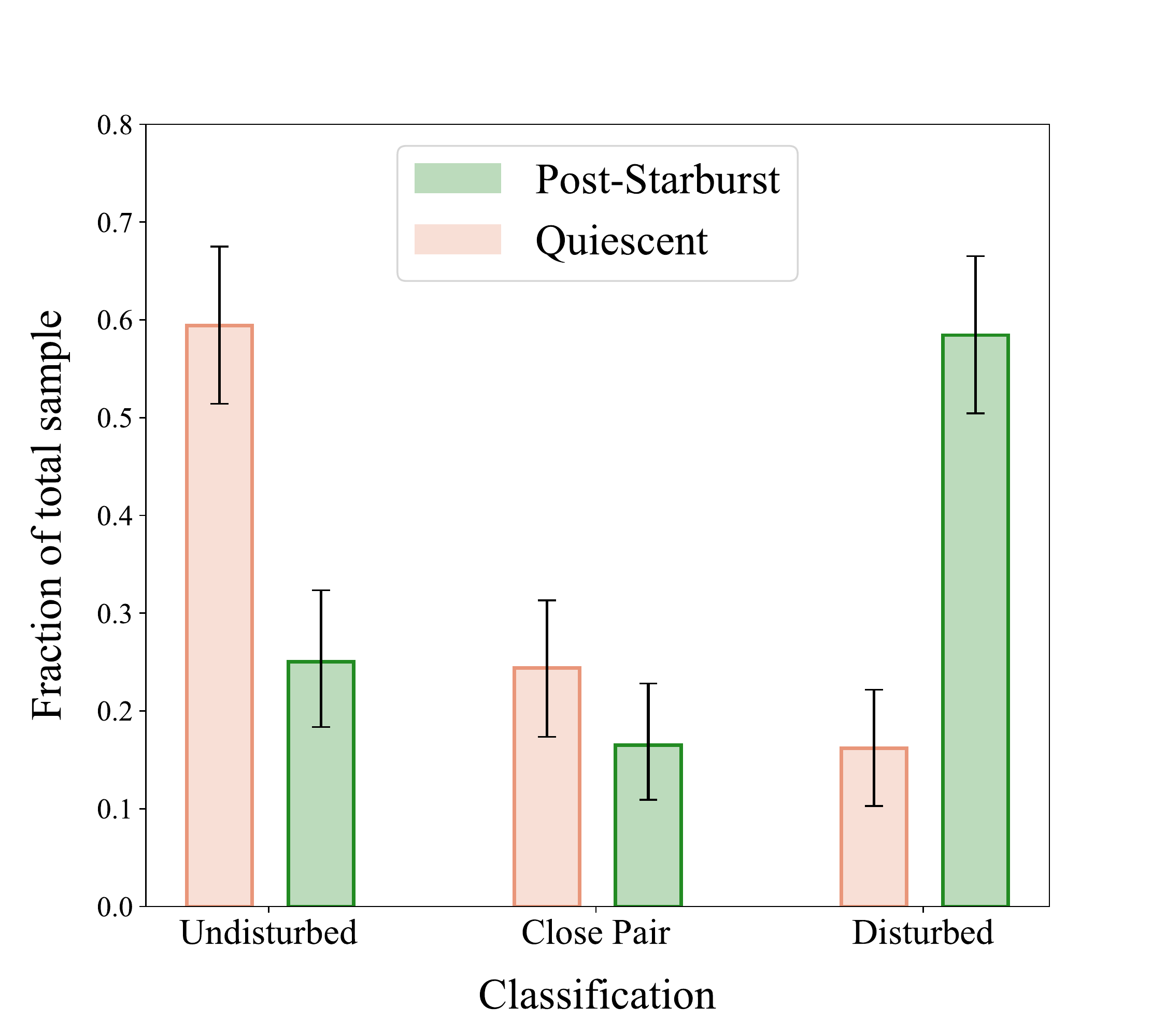}
\includegraphics[width=0.5\textwidth]{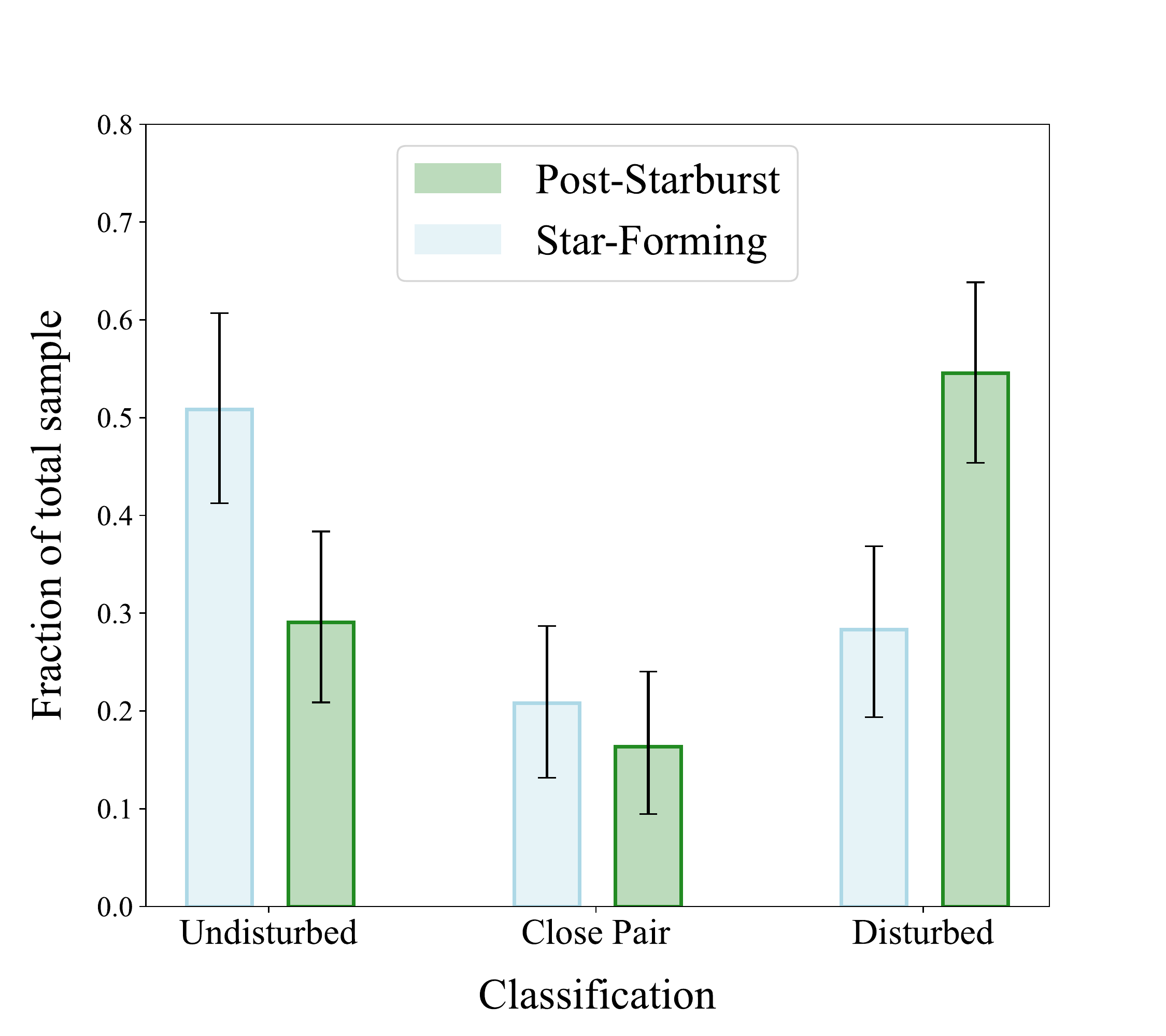}
\caption{Probability distribution functions for each of the samples of interest, with errorbars indicating 95\% confidence interval binomial errors.  (Right) Post-starburst versus quiescent galaxies.  Post-starburst galaxies are about four times as likely to be classified as disturbed when compared to quiescent galaxies.  (Left) Post-starburst versus star-forming galaxies.  Post-starburst galaxies are about twice as likely to be classified as disturbed when compared to star-forming galaxies.}
\label{fig:classification}
\end{figure*}

\subsection{Visual Classification of Tidal Disturbance}

Nine members of our team (MV, DJS, RB, JEG, KAS, ADG, JSS, VD, and GK) visually inspected our galaxy and comparison samples for asymmetry and major tidal disturbances.  Before classifying, classifiers participated in a group training and classifying session, in which we classified a sub-sample of 50 galaxies over all three samples 18 star-forming, 18 quiescent, and 17 post-starburst galaxies.  For both the group and individual classification, classifiers were given three images of most galaxies: two $i$-band images, one black on white and one white on black, with the stretch determined after masking the central pixels of the galaxy of interest to allow faint tidal features to appear despite the scaling, and one 3-color $riz$ image using the \texttt{Astropy} implementation of the \cite{Lupton2004} procedure for generating RGB images, with the \texttt{stretch} parameter set to 2 \citep{astropy2018}. Some galaxies did not have $r$ or $z$ band imaging, so classifiers were only given the two $i$-band images. An example galaxy image is shown in Fig.~\ref{fig:example}. To avoid bias, classifiers were not made aware of the original sample of each object. Classifiers flagged each galaxy as disturbed (2), close pair (1), or undisturbed (0). A "close pair" galaxy is a galaxy with a companion within 10 kpc in projected distance of the central galaxy.  We choose to flag these galaxies because pre-coalescence close pairs have been shown to correlate with enhancement of SFR \citep{Woods2010,patton2013} whereas post-mergers, rather than close pairs, show an enhancement in PSB fraction \citep{Goto2005b,Yamauchi2008,Ellison2022}. We note that because of the lack of spectroscopic confirmations of pairs, the close pair sample likely encompasses some chance encounters, but at projected distances of ~10 kpc we expect that most pairs should be associated for such massive galaxies \citep[see][]{Tal2013}.

 We consider the group classifications of each individual galaxy in the training sub-sample to be the ``correct" classification for the galaxy.  We compute the fraction of galaxies each reviewer identified correctly and weight each reviewer's overall classifications by this fraction. With this weight, we compute the weighted average fraction of galaxies which are classified in each category for each of the three samples.  In combination with the group training session, the weighted average mitigates the effect of any individual classifier who is not in agreement with the rest of the group.

Galleries of the full post-starburst, quiescent, and star-forming galaxy samples sorted by their classification scores are shown in Appendix~\ref{sec:appendix}.

\subsection{Results of Visual Classification}

\begin{figure}[t!]
\includegraphics[width=.45\textwidth]{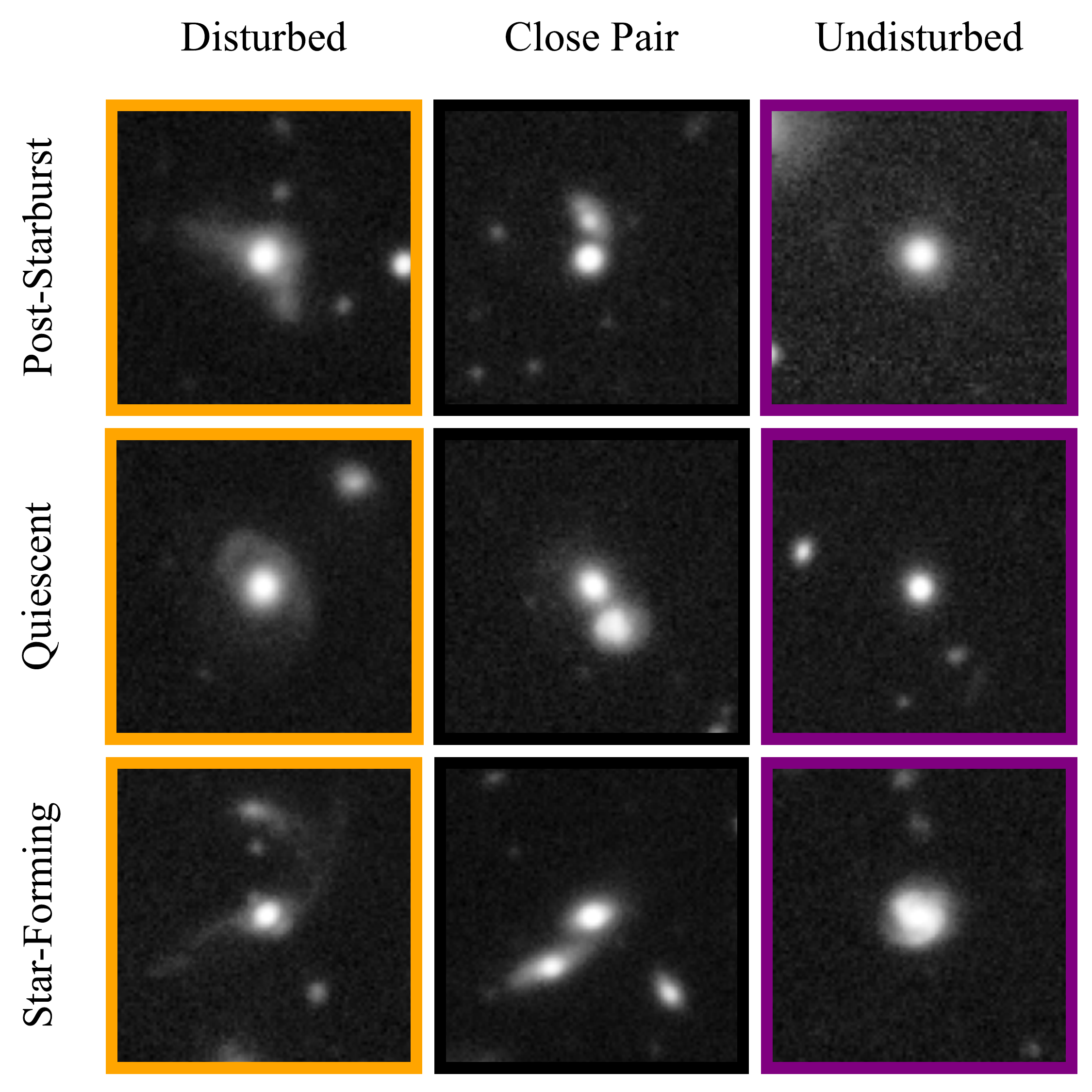}
\caption{12"x12" cutouts of our \squiggle post-starburst sample (top row) and our comparison LEGA-C samples (center row: quiescent, bottom row: star-forming). Galaxies are classified as either disturbed, close pair, or undisturbed.  For the galaxies above, either all or all but one of the classifiers agreed on the classification. Full galleries of the images of all 3 samples are shown in the appendix.
\label{fig:example_gallery_psb}}
\end{figure}

In Figure \ref{fig:classification} we show the weighted classification fractions among each of the three samples. Post-starburst galaxies are overall more disturbed than the comparison galaxies, especially the quiescent control galaxies. Post-starburst galaxies are classified as disturbed on-average about half of the time - $58^{+8.1}_{-8.0}\%$ percent of the time when compared to quiescent galaxies, and $58^{+9.3}_{-9.2}\%$ when compared to star-forming galaxies.  This is an excess of $3.6^{+2.9}_{-1.3}$ times the quiescent control sample and $2.1^{+1.9}_{-.73}$ times the star-forming sample.  As discussed above, it is also possible, though unlikely, that we have some starburst galaxies contaminating our control sample.  It is therefore possible that post-starburst galaxies are even more disturbed relative to star-forming galaxies than the numbers we report in this study. The small number of galaxies in the \squiggle survey with available HSC imaging means binomial error is dominant in this analysis.  The weighted scores fractions are not substantially different from the raw fractions; the systematic error from classifier selection is less than 0.3\% for all classification fractions. A representative sample of post-starburst, quiescent, and star-forming galaxies with associated classifications are shown in Figure \ref{fig:example_gallery_psb}. Galaxies outlined in orange are disturbed, galaxies outlined in purple are undisturbed, and the remaining galaxies are close pairs. Ambiguity may arise in visual classification, as tidal features and features like spiral arms, star-forming clumps, and satellite galaxies can be mistaken for one another.  For this reason, the classification of an individual galaxy may be unclear; however, the overall comparisons between our post-starburst and control samples show a clear excess of post-merger features in post-starburst galaxies over the controls.

\section{Discussion}

When compared to co-eval star-forming and quiescent galaxies, post-starbursts are more likely to be classified as disturbed. This finding is consistent with previous studies of low-redshift recently-quenched galaxies \citep{Zabludoff1996,Yang2008,Brown2009,Pracy2009,Pawlik2016,Sazonova2021}. If clear tidal features are the signatures of recent major mergers, the higher disturbance rate of post-starbursts is very suggestive of a causal link between galaxy mergers and galaxy quenching, at least at this late epoch.

\begin{figure}[hb!]
\includegraphics[width=0.48\textwidth]{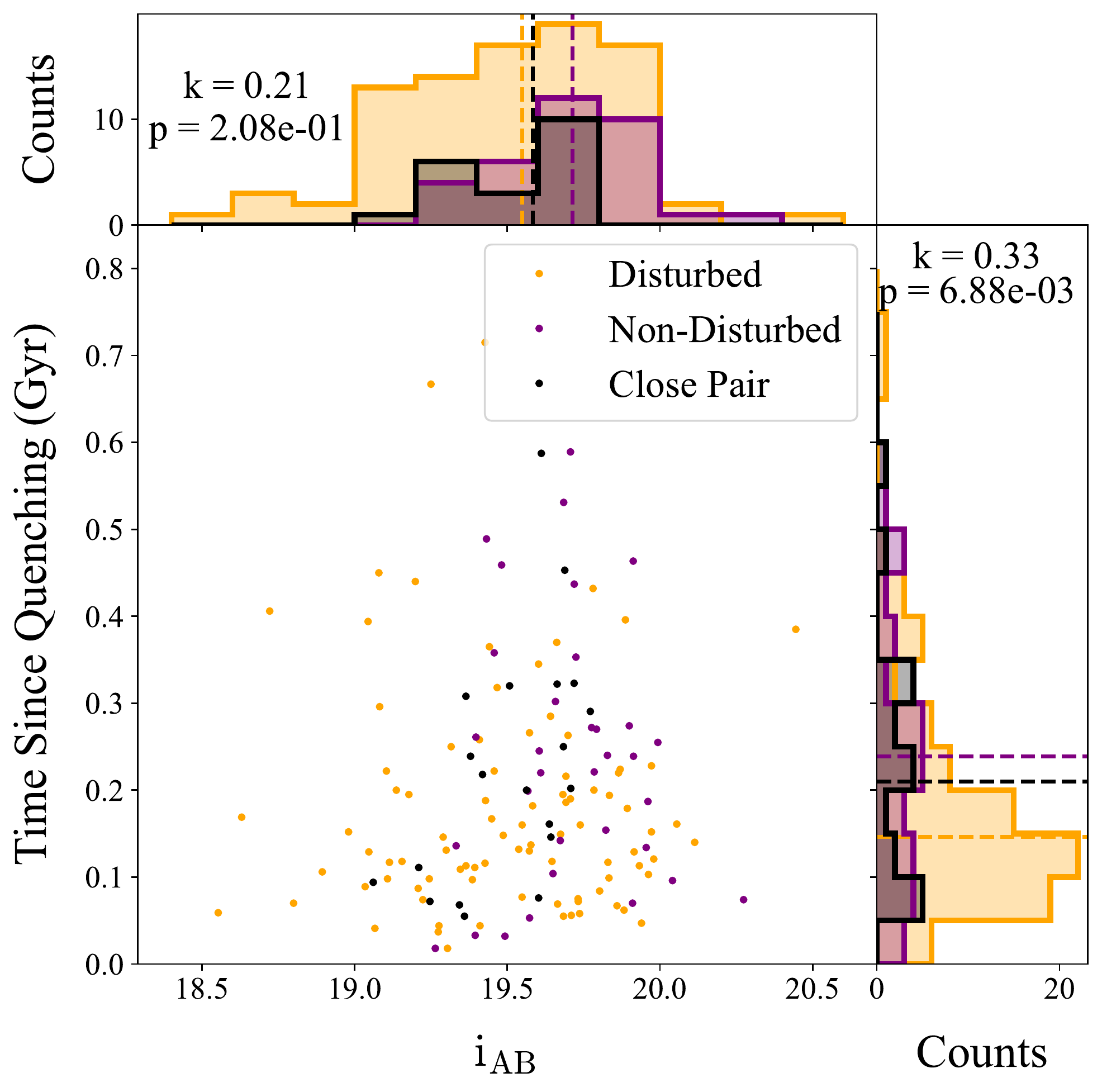}
\caption{The time since quenching (as measured from spectrophotometric fits in \citealt{Suess2022a}) versus the SDSS magnitude for the \squiggle sample of post-starburst galaxies, split into disturbed (orange), undisturbed (purple), and close pair (black) samples. Younger galaxies are more often classified as disturbed (two-sample Kolmogorov-Smirnov score of $k=0.33, p=6.9 \times 10^{-3}$), as are brighter galaxies ($k=0.21, p=2.1 \times 10^{-1}$). Although we cannot rule out some observational bias against identifying older or intrinsically fainter tidal features, both bright and young post-starburst galaxies exist without any merger signatures.
\label{fig:tq_imag}}
\end{figure}

In Figure \ref{fig:tq_imag}, we investigate the time since quenching, or the time since the end of the most recent starburst (see \citealt{Suess2022a, Suess2022b}), and SDSS $i$-band magnitudes of our disturbed and undisturbed samples to determine whether we are observing more than one path to quenching. The disturbed galaxies in our sample are, on the whole, younger and brighter than their undisturbed counterparts. The median time since quenching of the full sample is \tqmedian. We divide our sample into ``young" and ``old" galaxies for a time since quenching below or above the median, respectively. \youngdistfraction~of young galaxies and \olddistfraction~of old galaxies are disturbed, and the distribution in time since quenching is very different for disturbed and undisturbed galaxies in the sample. Simulations suggest that tidal features fade over time on a timescale of $\sim250$ Myr and have faded entirely within the first gigayear since quenching \citep{Pawlik2016, Sazonova2021}, implying some of the older undisturbed galaxies in our sample may still be major merger remnants for which tidal features have faded from view. However, 10 of the young ($\lesssim150$ Myrs post-quenching) galaxies in our sample appear to be undisturbed. Many physical factors of a galaxy merger can alter the brightness and scale of tidal features, including the mass and gas fractions of merging progenitor galaxies or the angle at which we view any given galaxy \citep{Lotz2008,Pawlik2016}. We therefore cannot rule out the possibility that these young undisturbed galaxies are also merger remnants but that we cannot see the tidal features associated with merging, either from physical or observational effects. Even so, it is also possible that these young undisturbed galaxies represent a non-merger pathway to rapid quenching.

\section{Conclusion}

We visually classify tidal disturbance in deep HSC imaging of 143 $\rm{M}_* \sim 10^{11} \rm{M}_\odot$ post-starburst galaxies at $z\sim0.7$ and find that $58^{+9.3}_{-9.2}\%$ of post-starburst galaxies show clear signs of tidal disturbance.  Post-starbursts are $3.6^{+2.9}_{-1.3}$ times more likely to be disturbed than comparison quiescent galaxies and $2.1^{+1.9}_{-.73}$ times more likely to be disturbed than star-forming galaxies. The disturbed galaxies are primarily bright and stopped forming stars within the last $\sim150$ Myrs. This bias is likely driven by two effects. First, it is easier to visually classify low surface brightness features in brighter galaxies. Second, tidal features fade on timescales of hundreds of Myrs, vanishing entirely about 700 Myrs after a starburst \citep{Conselice2006,Lotz2008,snyder2015a,pawlik2018,Sazonova2021}. It is therefore possible that more galaxies found in our sample have undergone a major merger but, due to the timescale on which tidal features fade and their low surface brightness, are flagged as undisturbed in this study. Our sample is universally more compact than co-eval star-forming and quiescent galaxies, which is also associated with gas-rich major mergers and may be a sign that this older population is indeed composed of merger remnants \citep{Setton2022}. However, the existence of 10 bright, recently-quenched, yet undisturbed galaxies suggests that a recent major merging event may not be universal in our sample. We conclude that major mergers are likely a significant factor in, but not the sole pathway to, galaxy quenching in massive post-starburst galaxies at intermediate redshift.

The need for additional pathways to quiescence in our sample leads us in several more speculative directions. Optical AGN are present in five of the \squiggle galaxies presented in this analysis and $5\pm0.7\%$ of the total \squiggle sample (compared to $1.5\pm0.4\%$ of the overall LEGA-C sample \citep{Barisic_2017} and likely play a role in galaxy quenching \citep{Greene2020}. The youngest post-starburst galaxies in our sample still host large molecular gas reservoirs \citep{Bezanson2022}, which can fuel pre-quenching star formation while simultaneously fueling a black hole. At the same time, our sample contains a galaxy with a large fraction of its cold gas in tidal tails, evidence that we see both cold gas reservoirs and the removal of cold gas in the \squiggle sample \citep{spilker2022}. The compactness of our sample may itself be tied to quenching, rather than just the remnant of major mergers \citep[e.g.][]{Franx2008}. Although likely not a rapid driver, the massive dark matter halos surrounding massive galaxies may play a role in halting the fueling of those galaxies \citep{Feldmann2015}, although this may be in tension with the remaining hydrogen gas present in our sample. Further study is needed to investigate these young undisturbed galaxies and their pathway to quenching.

Spectroscopic data are required to precisely measure the timescales of quenching. While we are currently using the best spectroscopic control sample of both star-forming and quiescent galaxies available to us, LEGA-C is not a perfect comparison.  Although we have matched in mass and redshift, we note that we have not accounted for systematic differences in, for example, the estimation of stellar masses.  Furthermore, the current samples are relatively small, and their statistical power is still somewhat limited. Characterizing the importance of merging in driving quenching in massive galaxies also requires asking similar questions when the process dominates at Cosmic Noon \citep{Whitaker2012a, Wild2016}. Future surveys, like those planned with the Subaru Prime Focus Spectrograph, will provide sufficiently large, wide-field data sets from which we will be able to select both post-starburst and mass-matched control star-forming and quiescent galaxies at higher and higher redshifts \citep{Takada2014}. These future data sets will be essential to understanding the process by which galaxies have transitioned from star formation to quiescence throughout the history of our universe. 

\acknowledgements 

MV acknowledges support from the NASA Pennsylvania Space Grant Consortium. DS, JEG, RB, and DN gratefully acknowledge support from NSF grant AST1907723. RB acknowledges support from the Research Corporation for Scientific Advancement (RCSA) Cottrell Scholar Award ID No: 27587. RF acknowledges financial support from the Swiss National Science Foundation (grant no 157591 and 194814). This work was performed in part at the Aspen Center for Physics, which is supported by National Science Foundation grant PHY-1607611.

The Hyper Suprime-Cam (HSC) collaboration includes the astronomical communities of Japan and Taiwan, and Princeton University. The HSC instrumentation and software were developed by the National Astronomical Observatory of Japan (NAOJ), the Kavli Institute for the Physics and Mathematics of the Universe (Kavli IPMU), the University of Tokyo, the High Energy
Accelerator Research Organization (KEK), the Academia Sinica Institute for Astronomy and Astrophysics in Taiwan (ASIAA), and Princeton University. 

Funding for SDSS-III has been provided by the Alfred P. Sloan Foundation, the Participating Institutions, the National Science Foundation, and the U.S. Department of Energy Office of Science. The SDSS-III web site is http://www.sdss3.org/. 

SDSS-III is managed by the Astrophysical Research Consortium for the Participating Institutions of the SDSS-III Collaboration including the University of Arizona, the Brazilian Participation Group, Brookhaven National Laboratory, Carnegie Mellon University, University of Florida, the French Participation Group, the German Participation Group, Harvard University, the Instituto de Astrofisica de Canarias, the Michigan State/Notre Dame/JINA Participation Group, Johns Hopkins University, Lawrence Berkeley National Laboratory, Max Planck Institute for Astrophysics, Max Planck Institute for Extraterrestrial Physics, New Mexico State University, New York University, Ohio State University, Pennsylvania State University, University of Portsmouth, Princeton University, the Spanish Participation Group, University of Tokyo, University of Utah, Vanderbilt University, University of Virginia, University of Washington, and Yale University.

%%%%%%%%%%%%%%%%%%%%%%%%%%%%%%%%%%%%%%%%%%%%%%%%%%%%%%%%%%%%%%%%%%%%%%%%%%%%%%%%%%%%%%%%%%%%%%%%%%%%%%%%%%%%%%%%%%%%%%%%%%%
%%%%%%%%%%%%%%%%%%%%%%%%%%%%%%%%%%%%%%%%%%%%%%%%%%%%%%%%%%%%%%%%%%%%%%%%%%%%%%%%%%%%%%%%%%%%%%%%%%%%%%%%%%%%%%%%%%%%%%%%%%%

\newpage

\bibliography{SQuIGGLE_Mergers}

\appendix

\section{Full Image Galleries}\label{sec:appendix}

% The following image galleries are also available on the \squiggle website at \url{https://squigglesurvey.github.io/}.

\begin{figure}[!h]
\includegraphics[width=0.95\textwidth]{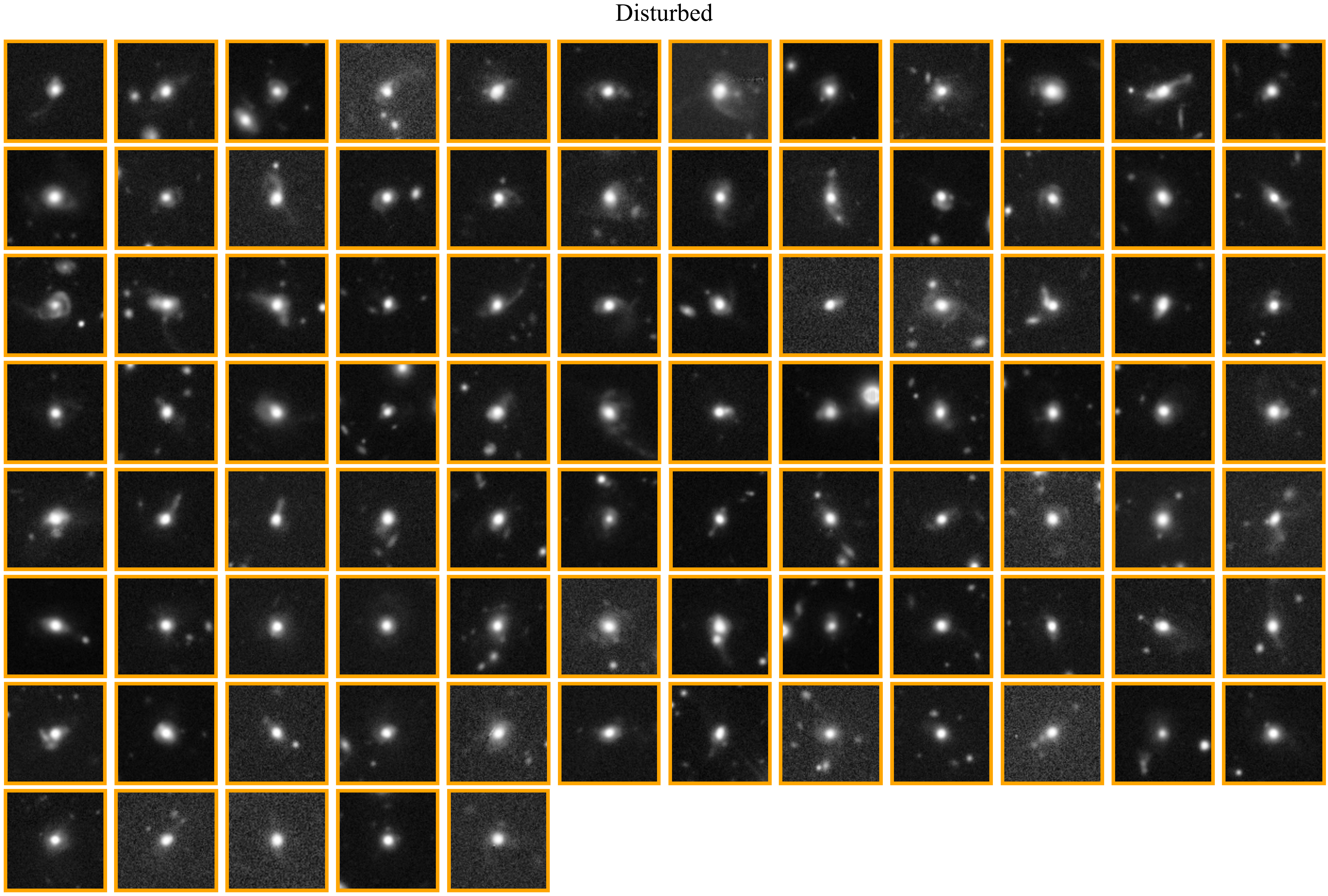}
\includegraphics[width=0.95\textwidth]{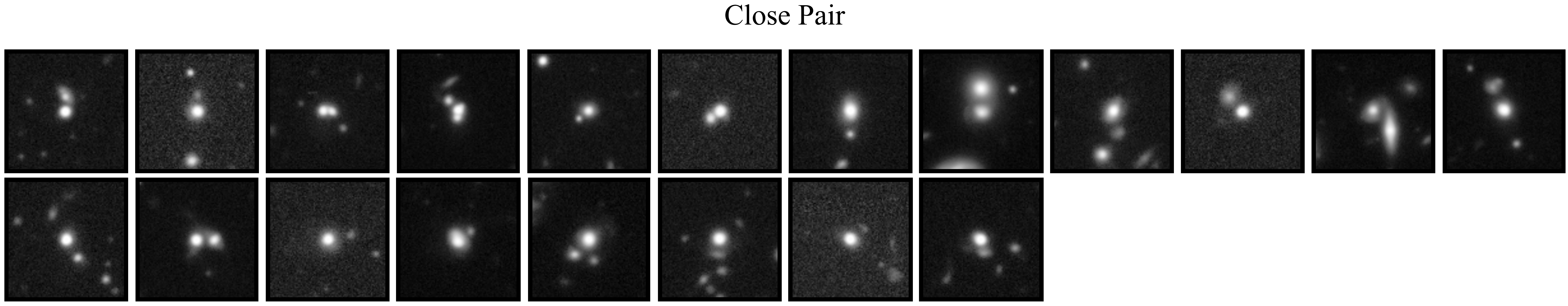}
\includegraphics[width=0.95\textwidth]{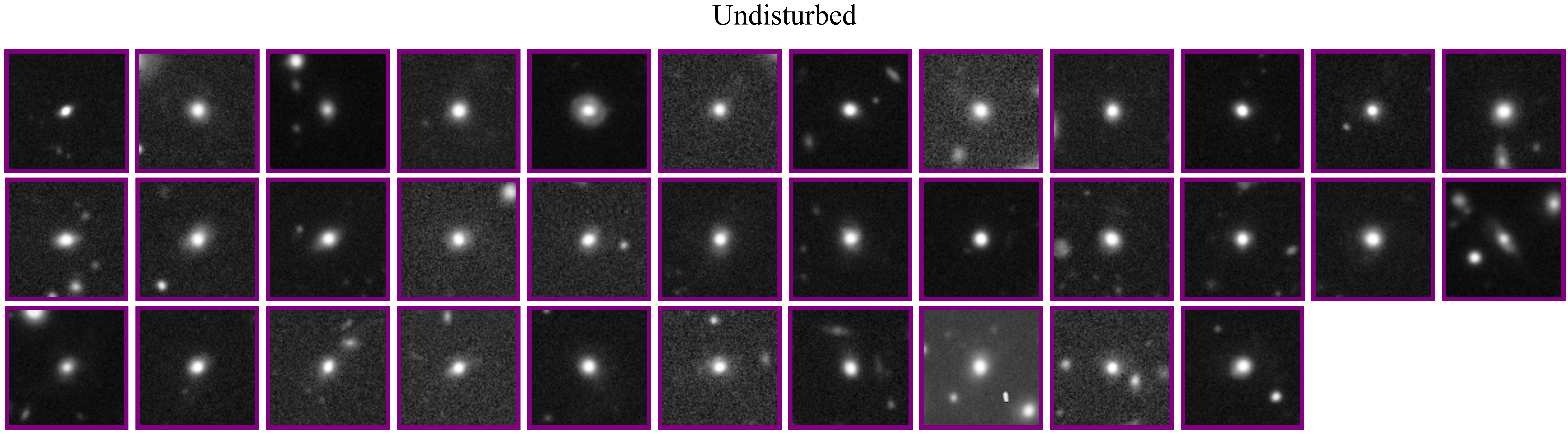}
\caption{The full sample of \squiggle post-starburst galaxies sorted into disturbed, close pair, and undisturbed galaxies.  Galaxies bordered in orange are classified as disturbed, galaxies bordered in black are close pairs, and galaxies bordered in orange are undisturbed.  
\label{fig:psb_gallery}}
\end{figure}

\begin{figure}
\includegraphics[width=0.95\textwidth]{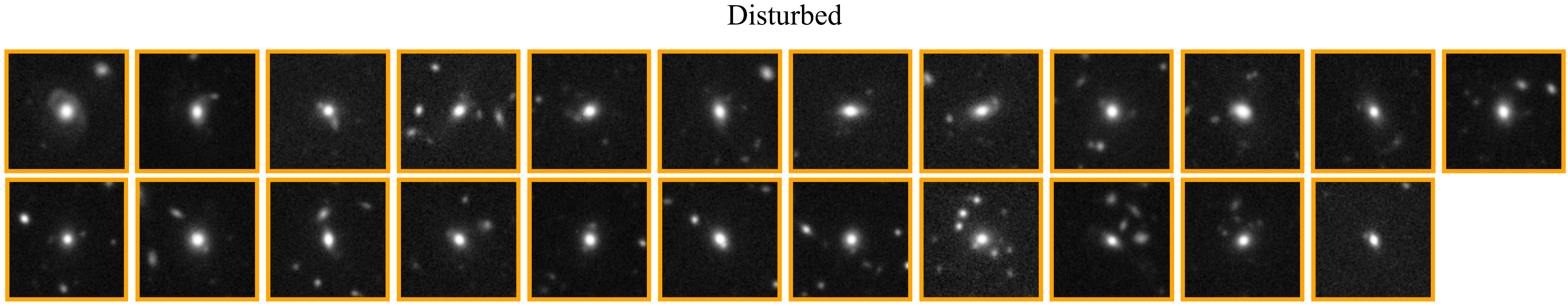}
\includegraphics[width=0.95\textwidth]{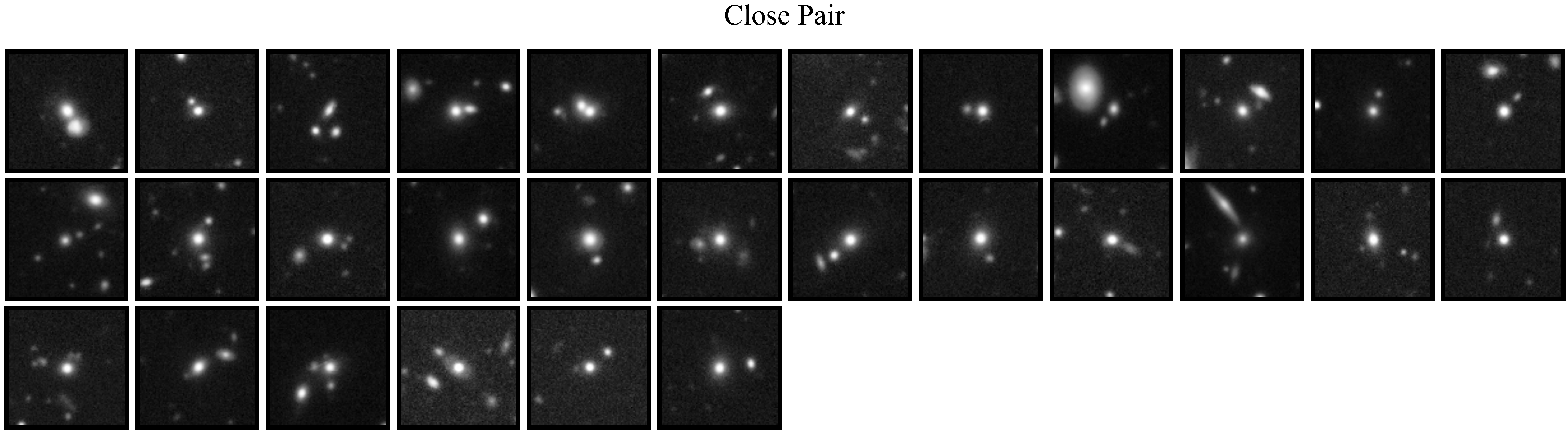}
\includegraphics[width=0.95\textwidth]{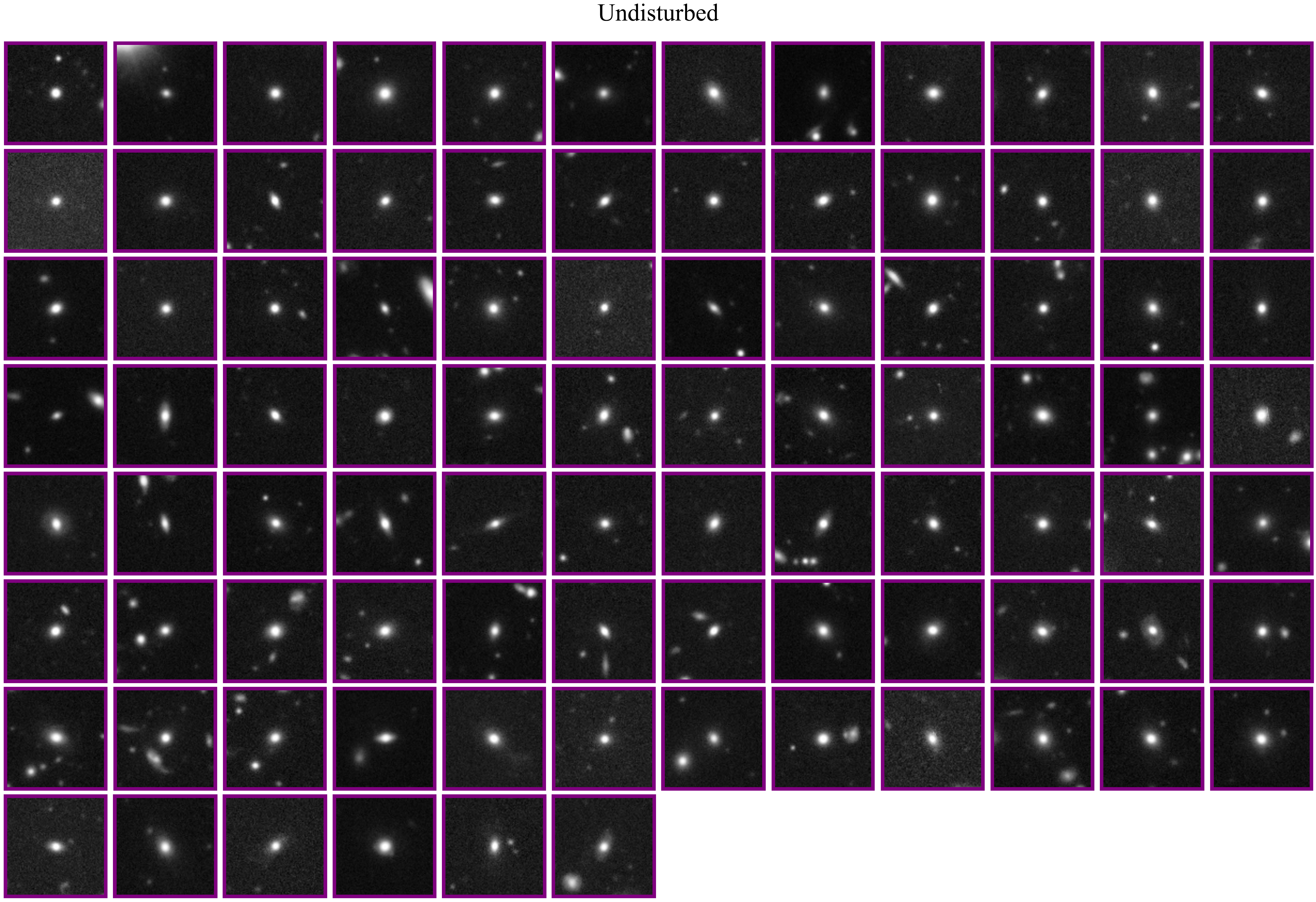}
\caption{The full sample of LEGA-C quiescent comparison galaxies sorted into disturbed, close pair, and undisturbed galaxies.  Galaxies bordered in orange are classified as disturbed, galaxies bordered in black are close pairs, and galaxies bordered in orange are undisturbed.  
\label{fig:q_gallery}}
\end{figure}

\begin{figure}
\includegraphics[width=0.95\textwidth]{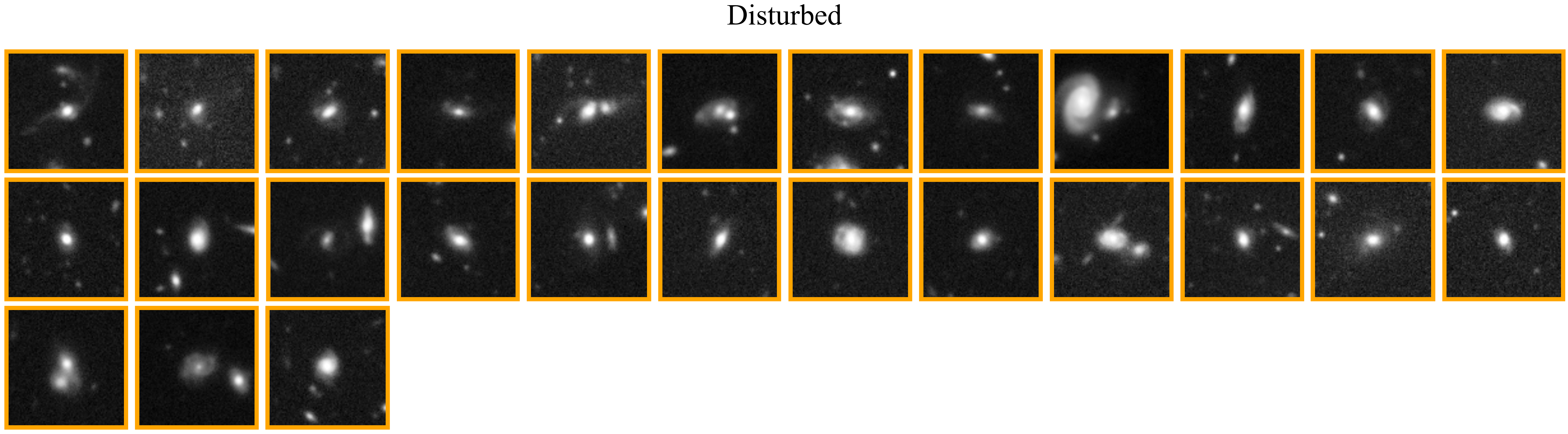}
\includegraphics[width=0.95\textwidth]{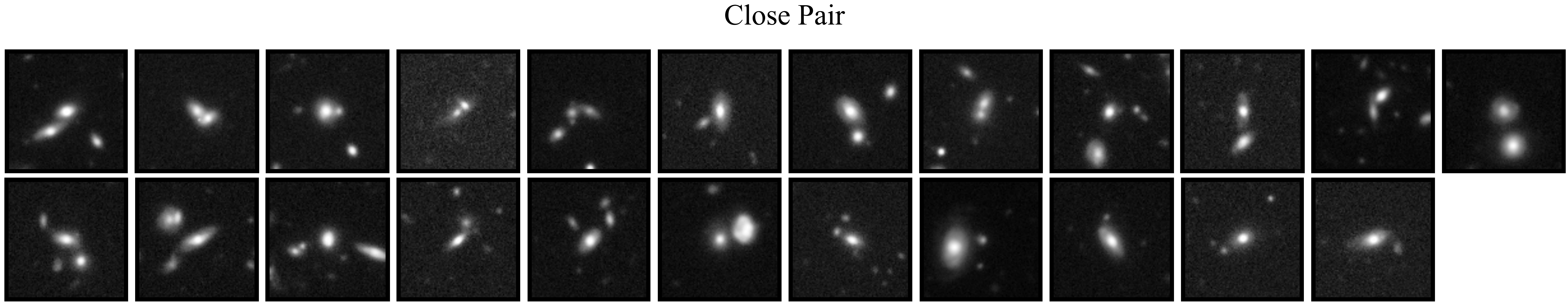}
\includegraphics[width=0.95\textwidth]{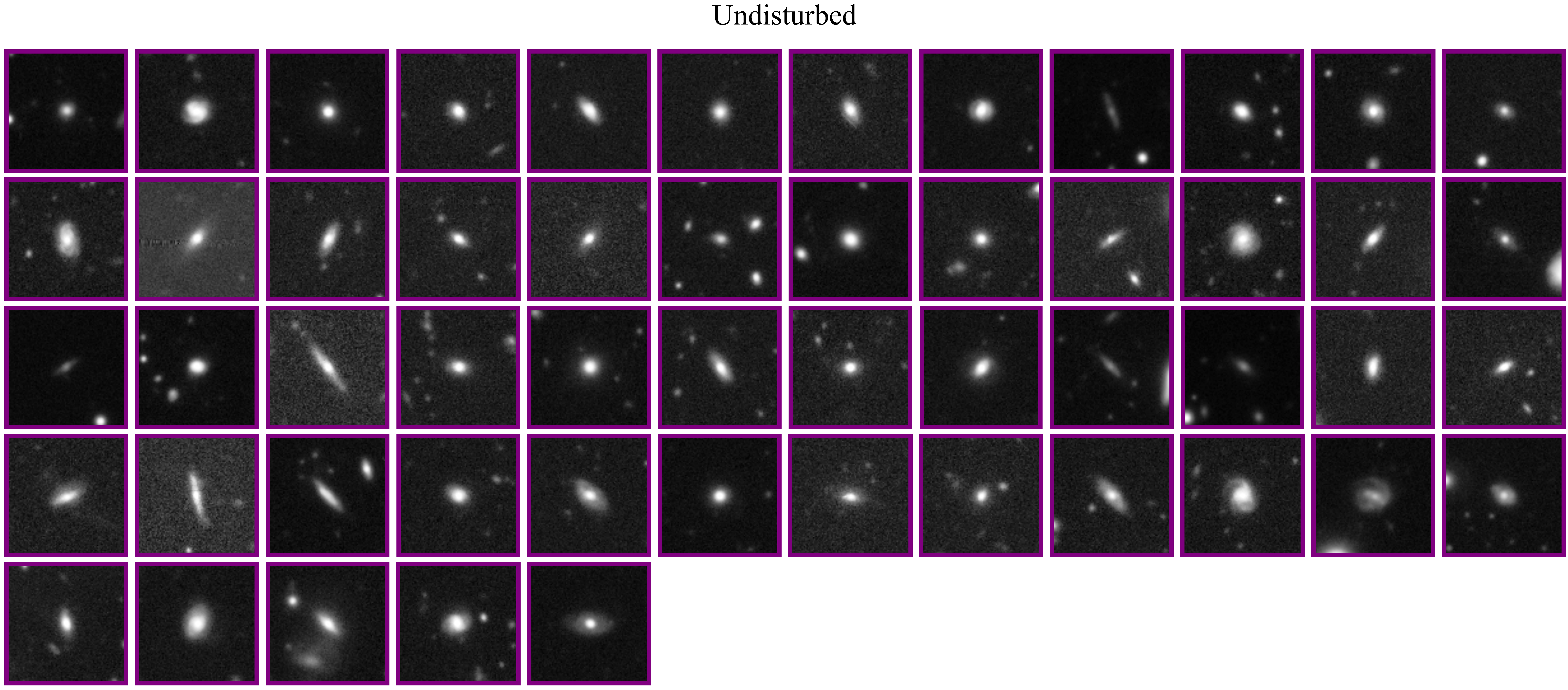}
\caption{The full sample of LEGA-C star-forming comparison galaxies sorted into disturbed, close pair, and undisturbed galaxies.  Galaxies bordered in orange are classified as disturbed, galaxies bordered in black are close pairs, and galaxies bordered in orange are undisturbed.  
\label{fig:sf_gallery}}
\end{figure}

Here, we present galleries of the \squiggle post-starburst (Figure \ref{fig:psb_gallery}), LEGA-C quiescent (Figure \ref{fig:q_gallery}), and LEGA-C star-forming (Figure \ref{fig:sf_gallery}) galaxy cutouts for the entirety of each sample. The images are on the same 12"$\times$12" scale as in Figure \ref{fig:example_gallery_psb} and are sorted by their classification as disturbed, close pair, or undisturbed.  Classification for an individual galaxy is determined by the most common classification given by our team of nine classifiers.  As such, the fractions of disturbed, close pair, and undisturbed galaxies in these figures are not the same as the weighted fractions quoted in this paper.

\end{document}